\documentclass[
    ,final            
  ]
  {aipproc}
\layoutstyle{6x9}
\newcommand{\lsim}{\stackrel{<}{_\sim}}

\newcommand{\tppp}{\tau\to\pi\pi\pi\nu_\tau}
\newcommand{\tpppc}{\tau^-\to\pi^+\pi^-\pi^-\nu_\tau}
\newcommand{\tpppn}{\tau^-\to\pi^-\pi^0\pi^0\nu_\tau}

\begin{document}

\title{Hadronic decays of the tau lepton~: 
$\tau^- \rightarrow \left(\pi \pi \pi \right)^- \nu_{\tau}$ within
Resonance Chiral Theory \footnote{Talk given by J.~P. 
at the International Workshop
on Quantum Chromodynamics: Theory and Experiment (QCD@Work 2005), 
Conversano (Bari, Italy), 16th-20th June 2005; Reports:
IFIC/05-41, FTUV/05-0920. To appear in the Proceedings.} }


\author{D.~G\'omez Dumm}{address={IFLP, CONICET - Depto. de F\'{\i}sica, 
Univ. Nac. de la Plata, C.C. 67, 1900 La Plata, Argentina}
}
\author{A.~Pich}{
  address={Instituto de F\'{\i}sica Corpuscular, IFIC, CSIC-Universitat
  de Val\`encia, Edifici d'Instituts de Paterna, Apt. Correus 22085,
  E-46071 Val\`encia, Spain}
}
\author{J.~Portol\'es}{
  address={Instituto de F\'{\i}sica Corpuscular, IFIC, CSIC-Universitat
  de Val\`encia, Edifici d'Instituts de Paterna, Apt. Correus 22085,
  E-46071 Val\`encia, Spain}
}

\begin{abstract}
 $\tau$ decays into hadrons foresee the study of the hadronization
 of vector and axial-vector QCD currents, yielding relevant information
 on the dynamics of the resonances entering into the processes. We
 analyse $\tau \rightarrow \pi \pi \pi \nu_{\tau}$ decays within 
 the framework of the Resonance Chiral Theory, comparing this theoretical
 scheme with the experimental data, namely ALEPH spectral function and
 branching ratio. Hence we get values for the mass and on-shell width
 of the $a_1(1260)$ resonance, and provide the structure functions
 that have been measured by OPAL and CLEO-II. 
\end{abstract}

\maketitle


\section{Introduction}

The implementation of Quantum Chromodynamics (QCD)
in the energy region populated by light-flavoured resonances
($M_{\rho} \lsim E \lsim 2 \; \mbox{GeV}$, being $M_{\rho}$ the 
mass of the $\rho(770)$) is a demanding task that involves
poorly known aspects such as bound and resonant states, duality
and hadronization mechanisms. Though {\em ad hoc }
Breit-Wigner parameterisations
have been widely employed in the literature \cite{Pich:1989pq,Achasov:1996gw}
they are not necessarily consistent with the underlying theory,
as they seem to violate the chiral symmetry of massless QCD 
\cite{Portoles:2000sr,GomezDumm:2003ku}.
\par
$\tau$ decays into hadrons allow to study the hadronization properties
of vector and axial-vector QCD currents and, accordingly, to determine
intrinsic properties of the dynamics generated by resonances
\cite{Portoles:2004vr}. 
At very low energies, typically $E \ll M_{\rho}$, Chiral Perturbation
Theory ($\chi$PT) \cite{Weinberg:1978kz} is the Effective Field Theory
of QCD. Still the $\tau \rightarrow \pi \pi \pi \nu_{\tau}$ decays,
through their full energy spectrum, are driven by the $\rho (770)$ and
$a_1(1260)$ resonances mainly, in an energy region where the invariant
hadron momentum approaches the masses of the resonances. Hence
$\chi$PT is no longer applicable to the study of the whole spectrum but
only to the very low energy domain \cite{Colangelo:1996hs}. The standard
procedure that has been followed \cite{Pich:1989pq} to deal with these
decays has been to modulate the amplitudes with a Breit-Wigner
parameterization, fixing the normalization in order to match the
leading ${\cal O}(p^2)$ $\chi$PT. Nevertheless, its deviation of the
chiral behaviour
at higher orders \cite{Portoles:2000sr,GomezDumm:2003ku} could spoil
any outcome provided by the analysis of data.
\par
Lately several experiments have collected good quality data on 
$\tau \rightarrow \pi \pi \pi \nu_{\tau}$, such as branching ratios
and spectra \cite{Barate:1998uf} or structure functions 
\cite{Ackerstaff:1997dv}. Their analysis within a model-independent
framework is highly desirable if one wishes to collect information
on the hadronization of the relevant QCD currents.

\section{The Resonance Chiral Theory of QCD}

At energies $E\sim M_{\rho}$ the resonance mesons are active degrees
of freedom that have to be properly included into the pertinent Lagrangian.
The procedure, put forward in Refs.~\cite{Ecker:1988te,Ecker:1989yg}
and known as Resonance Chiral Theory (R$\chi$T), is ruled by the 
approximate chiral symmetry of QCD, that drives the interaction of 
Goldstone bosons (the lightest octet of pseudoscalar mesons), and 
the $SU(3)_V$ assignments of the resonance multiplets. This construction
is embedded within a comprehensive framework guided by the 
large number of colours ($N_C$) limit of QCD \cite{'tHooft:1974hx}. 
The $1/N_C$ expansion tells us that, at leading order, we should only
consider the tree level diagrams given by a local Lagrangian with infinite
zero-width states in the spectrum. This is precisely the role of R$\chi$T.
However in most processes, like hadron tau decays, we need to include
finite widths that only appear at next-to-leading order in the large-$N_C$
expansion and, moreover, we will only include one multiplet of vector and 
axial-vector resonances in our theory. Thus, in practice, we have to 
model this large-$N_C$ expansion to some extent.
\par
The final hadron system in the $\tau \rightarrow \pi \pi \pi \nu_{\tau}$
decays spans a wide energy region, namely 
$3 \, m_{\pi} \lsim E \lsim M_{\tau}$ that is populated by many resonances.
R$\chi$T is the appropriate framework to work with and we consider the
Lagrangian \cite{GomezDumm:2003ku,Ecker:1988te}~:
\begin{eqnarray}
\label{eq:ret}
{\cal L}_{\rm R\chi T}   & =   &  
\frac{F^2}{4}\langle u_{\mu}
u^{\mu} + \chi _+ \rangle \, + \, \frac{F_V}{2\sqrt{2}} \langle V_{\mu\nu}
f_+^{\mu\nu}\rangle \,
+ \, i \,\frac{G_V}{\sqrt{2}} \langle V_{\mu\nu} u^\mu
u^\nu\rangle  \, + \, 
\frac{F_A}{2\sqrt{2}} \langle A_{\mu\nu}
f_-^{\mu\nu}\rangle \,\nonumber \\
& &  + \, {\cal L}_{\rm kin}^{\rm V} \, + \,  {\cal L}_{\rm kin}^{\rm A} \, + 
\, \sum_{i=1}^{5}  \, \lambda_i  \,
{\cal O}^i_{\rm VAP} \, ,
\end{eqnarray}
where all the couplings are real, being $F$ the decay constant of the pion
in the chiral limit, and the operators ${\cal O}_{VAP}^i$
are given by~:
\begin{eqnarray}
\label{eq:lag2}
{\cal O}^1_{\rm VAP} &  = & \langle \,  [ \, V^{\mu\nu} \, , \, 
A_{\mu\nu} \, ] \,  \chi_- \, \rangle \; \; , \nonumber \\
{\cal O}^2_{\rm VAP} & = & i\,\langle \, [ \, V^{\mu\nu} \, , \, 
A_{\nu\alpha} \, ] \, h_\mu^{\;\alpha} \, \rangle \; \; , \\
{\cal O}^3_{\rm VAP} & = &  i \,\langle \, [ \, \nabla^\mu V_{\mu\nu} \, , \, 
A^{\nu\alpha}\, ] \, u_\alpha \, \rangle \; \; ,  \nonumber \\
{\cal O}^4_{\rm VAP} & = & i\,\langle \, [ \, \nabla^\alpha V_{\mu\nu} \, , \, 
A_\alpha^{\;\nu} \, ] \,  u^\mu \, \rangle \; \; , \nonumber \\
{\cal O}^5_{\rm VAP} & =  & i \,\langle \, [ \, \nabla^\alpha V_{\mu\nu} \, , \, 
A^{\mu\nu} \, ] \, u_\alpha \, \rangle \nonumber \; \; .
\end{eqnarray}
The notation is that of Ref.~\cite{Ecker:1988te}. Notice that we are 
using the antisymmetric tensor formulation to describe the spin $1$
resonances and, consequently, we do not consider the ${\cal O}(p^4)$
$\chi$PT Lagrangian of Goldstone bosons \cite{Ecker:1989yg}.
\par
Our Lagrangian theory has eight a priori unknown coupling constants, 
namely $F_V$, $F_A$, $G_V$ and $\lambda_i$, $i=1,...5$. Phenomenology
could provide direct information on them; for instance $F_V$ could be
extracted from the measured $\Gamma(\rho^0 \rightarrow e^+e^-)$, 
$G_V$ from $\Gamma(\rho^0 \rightarrow \pi^+ \pi^-)$, $F_A$ from
$\Gamma(a_1 \rightarrow \pi \gamma)$ and the $\lambda_i$ appear in 
$\Gamma(a_1 \rightarrow \rho \pi)$ or the $\tppp$ processes themselves.
It is conspicuous, though, that ${\cal L}_{R \chi T}$ is not QCD for
arbitrary values of the couplings. Hence if we want to comprehend more about
QCD in this non-perturbative regime we should try to learn about the
determination of the couplings from the underlying theory
\cite{Pich:2002xy}. On this account we will implement several known
features of the strong interaction theory in the following.
\par
The QCD ruled short--distance behaviour of the vector and axial-vector form 
factors in the large--$N_C$ limit (approximated with only one octet
of vector resonances) constrains the couplings of 
${\cal L}_{\rm R\chi T}$ in Eq.~(\ref{eq:ret}), which must
satisfy \cite{Ecker:1989yg}~:
\begin{eqnarray}
\label{fvgv}
1 \, - \, \frac{F_V \, G_V}{F^2} & = & 0 \; \; , \nonumber \\
2 F_V G_V \, - \, F_V^2 \, & = & 0 \; \; .
\end{eqnarray} 
In addition, the first Weinberg sum rule, in the limit 
where only the lowest narrow resonances contribute to the vector 
and axial--vector spectral functions, leads to
\begin{equation}
\label{fvfa}
F_V^2 - F_A^2 = F^2 \;.
\end{equation}
In this way all three couplings $F_V$, $G_V$ and $F_A$
can be written in terms of the pion decay constant~: 
$F_V = \sqrt{2} F$, $G_V = F/\sqrt{2}$ and $F_A = F$. These results are
well satisfied phenomenologically and we have adopted them. In the
next section we will comment on an analogous study of the $\lambda_i$
couplings.

\section{The axial-vector form factors 
in $\tau^- \rightarrow \left(\pi \pi \pi \right)^- \nu_{\tau}$}

The decay amplitudes for the 
$\tau^- \rightarrow \pi^+ \pi^- \pi^-  \nu_{\tau}$ and
$\tau^- \rightarrow \pi^- \pi^0 \pi^0  \nu_{\tau}$ processes can be
written as
\begin{equation} 
{\cal M}_{\pm}  \, = \,  - \,
\frac{G_F}{\sqrt{2}} \, V_{ud} \, \bar u_{\nu_\tau}
\gamma^\mu\,(1-\gamma_5) u_\tau\, T_{\pm \mu} \; , 
\end{equation}
\begin{equation} \label{eq:tmu1}
T_{\pm \mu}(p_1,p_2,p_3) \,  =  \,  
 \langle  \pi_1(p_1)\pi_2(p_2)\pi^{\pm}(p_3)  |   {\cal A}_\mu  |  
 0  \rangle \, ,
\end{equation}
as in the isospin limit there is no contribution of the vector current
to these processes. In $T_{\pm \mu}(p_1,p_2,p_3)$ the
$\pi^+$ is the one in $\tpppc$ and
$\pi^-$ that in $\tpppn$. The hadronic tensor can be written in terms of three
form factors, $F_1$, $F_2$ and $F_P$, as \cite{Kuhn:1992nz}~:
\begin{equation}
T^{\mu} \; = \; V_1^\mu\,F_1 \, + \,  V_2^\mu\,F_2 \, + \,  
Q^\mu\,F_P \; \; ,
\label{tmu}
\end{equation}
where
\begin{eqnarray}
V_1^\mu & = & \left( \, g^{\mu \nu} \, - \, 
\frac{Q^{\mu} Q^{\nu}}{Q^2} \, \right) \, ( \, p_1 - p_3 \, )_{\nu} \; \; ,
\nonumber \\
V_2^\mu & = & \left( \, g^{\mu \nu} \, - \, 
\frac{Q^{\mu} Q^{\nu}}{Q^2} \, \right) \, ( \, p_2 - p_3 \, )_{\nu} \; \; ,
\nonumber \\
Q^\mu & = &  p_1^\mu + p_2^\mu + p_3^\mu\; \; .
\end{eqnarray}
The form factors $F_1$ and 
$F_2$ have a transverse structure in the 
total hadron momenta $Q_{\mu}$ and drive a
$J^P=1^+$ transition. Bose symmetry under interchange of the
two identical pions in the final state demands that 
$F_1(Q^2,s,t) = F_2(Q^2,t,s)$ where $s=(p_1+p_3)^2$ and $t=(p_2+p_3)^2$.
Meanwhile $F_P$ accounts for a $J^P=0^-$
transition that carries pseudoscalar degrees of 
freedom and vanishes with the square of the pion mass. Its contribution
to the spectral function of $\tppp$ goes like $m_{\pi}^4/Q^4$ and, accordingly,
it is very much suppressed with respect to that coming from $F_1$ and $F_2$.
We will not consider it in the following. 
\par
\begin{figure}
\includegraphics*[scale=0.99,clip]{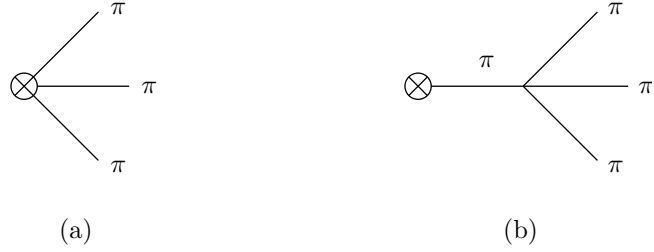}
\caption{\label{fig:diag1}Diagrams contributing to the hadronic
amplitude  $T_{\pm \mu}$ at ${\cal O}(p^2)$ $\chi$PT. }
\end{figure}
\begin{figure}
\includegraphics*[scale=0.99,clip]{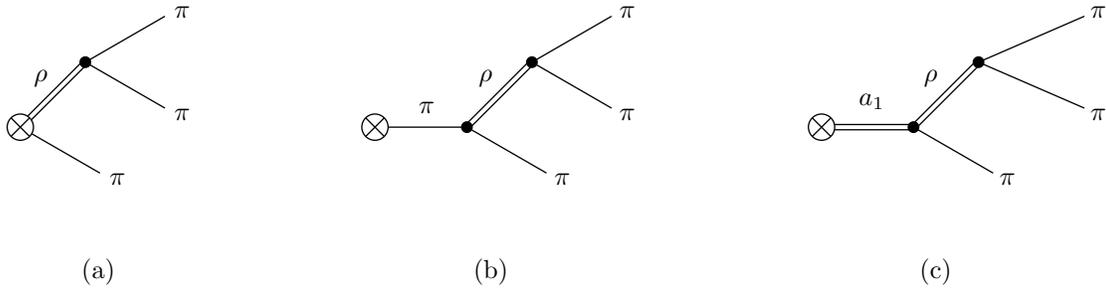} 
\caption{\label{fig:diag2} Resonance--mediated diagrams contributing
to $T_{\pm \mu}$.}
\end{figure}

In the low $Q^2$ region, the matrix element in Eq.~(\ref{eq:tmu1}) can be
calculated using $\chi$PT. At ${\cal O}(p^2)$ one has two contributions, 
arising from the diagrams in Fig.~\ref{fig:diag1}. 
The sum of both graphs yields
\begin{equation}
T_{\pm\mu}^{\chi} \, = \, 
 \mp \frac{2\sqrt{2}}{3 F}\left\{ V_{1\mu} + V_{2\mu} \right\}  \; \; .
\label{lowq}
\end{equation}
We now include the resonance--mediated contributions
to the amplitude, to be evaluated through the interacting terms
in ${\cal L}_{\rm R\chi T}$ Eq.~(\ref{eq:ret}). 
The relevant diagrams to be taken into account are those shown in 
Fig.~\ref{fig:diag2}. We get 
\begin{eqnarray}
\label{eq:t1r}
T_{\pm\mu}^{R} & = & \mp \frac{\sqrt{2}\,F_V\,G_V}{3\,F^3} \; 
\left[ \alpha(Q^2,s,t) \, V_{1\mu} + \alpha(Q^2,t,s) \, V_{2\mu} \right]
 \nonumber \\
 & & \pm \frac{4 \, F_A \, G_V}{3 \,F^3} \, 
 \frac{Q^2}{Q^2-M_A^2} \, 
 \left[ \beta(Q^2,s,t) \, V_{1 \mu} + \beta(Q^2,t,s) \, V_{2 \mu}
 \right] \; . 
\end{eqnarray}
The functions $\alpha(Q^2,s,t)$
and $\beta(Q^2,s,t)$ are~:
\begin{equation}
\alpha(Q^2,s,t) \, = \,   - \, 3 \,  \frac{s}{s-M_V^2} \, + \, 
\left( \frac{2 G_V}{F_V} - 1 \right) \, \left\lbrace 
\, \frac{2 Q^2-2s-u}{s-M_V^2} \, + \, \frac{u-s}{t-M_V^2} \, \right\rbrace 
\;\;, 
\end{equation}
\begin{equation}
\beta(Q^2,s,t) \, = \, 
- \, 3 \, (\lambda' + \lambda'') \, \frac{s}{s-M_V^2} \, 
+ \, \, F(Q^2,s) \, \frac{2 Q^2 + s - u}{s-M_V^2} \, 
+ \,  F(Q^2,t) \, \frac{u-s}{t-M_V^2} \, \; \; ,
\label{ff2r}
\end{equation}
with
\begin{equation} \label{eq:fq2}
F(Q^2,s)  =  - \,\lambda_0\, \frac{m_\pi^2}{Q^2} \, +  \, 
\lambda'\, \frac{s}{Q^2} \, + \,  \lambda''  \; , 
\end{equation}
and depend
on three combinations of the $\lambda_i$ couplings in 
${\cal L}_{\rm R \chi T}$, that we call $\lambda_0$, $\lambda'$ and 
$\lambda''$. Following the ideas outlined above
we can get information on these combinations
by implementing known aspects of asymptotic QCD. In particular 
we expect that the form factor of the axial-vector current into
three pions should vanish at infinite transfer of momentum 
($Q^2 \rightarrow \infty$). This is a consequence of the fact that
its contribution to the spectral function of the axial-vector current
correlator, being positive, has to add to other infinite hadronic
positive contributions to reach the constant value evaluated within
QCD \cite{Floratos:1978jb}.
Accordingly the proper behaviour of the 
$T_{\pm \mu}$ form factor imposes the constraints~:
\begin{eqnarray}
2 \, \lambda' \, - \, 1 & = & 0 \; \;, \nonumber \\
\lambda'' & = & 0 \; \;.
\label{eq:rellam}
\end{eqnarray}
Hence there is only one combination of couplings left unknown,
namely $\lambda_0$.
\par
Finally an additional comment on the result for 
$T_{\pm \mu}^R$ is required. The form factors in Eq.~(\ref{eq:t1r}) include 
zero--width $\rho$(770) and
$a_1$(1260) propagator poles, leading to divergent phase--space integrals 
in the calculation of the $\tppp$ decay width as the kinematical variables
go along the full energy spectrum. The result can be regularized
through the inclusion of resonance widths, which means to go beyond the
leading order in the $1/N_C$ expansion, and implies the introduction of
some additional theoretical inputs. This issue has been analysed in detail 
within the resonance chiral effective theory in 
Ref.~\cite{GomezDumm:2000fz} and, 
accordingly, we include off--shell widths for both resonances 
\cite{GomezDumm:2003ku}.

\section{Theory versus Experiment}
To analyse the experimental data we will only consider the dominating
$J^P = 1^+$ driven axial--vector form factors, that satisfy
$T_{+\mu} = - T_{-\mu}$ hence providing the same predictions for both
$\tpppc$ and $\tpppn$ processes in the isospin limit.
\par
\begin{figure}
\includegraphics*[angle=-90,scale=0.6,clip]{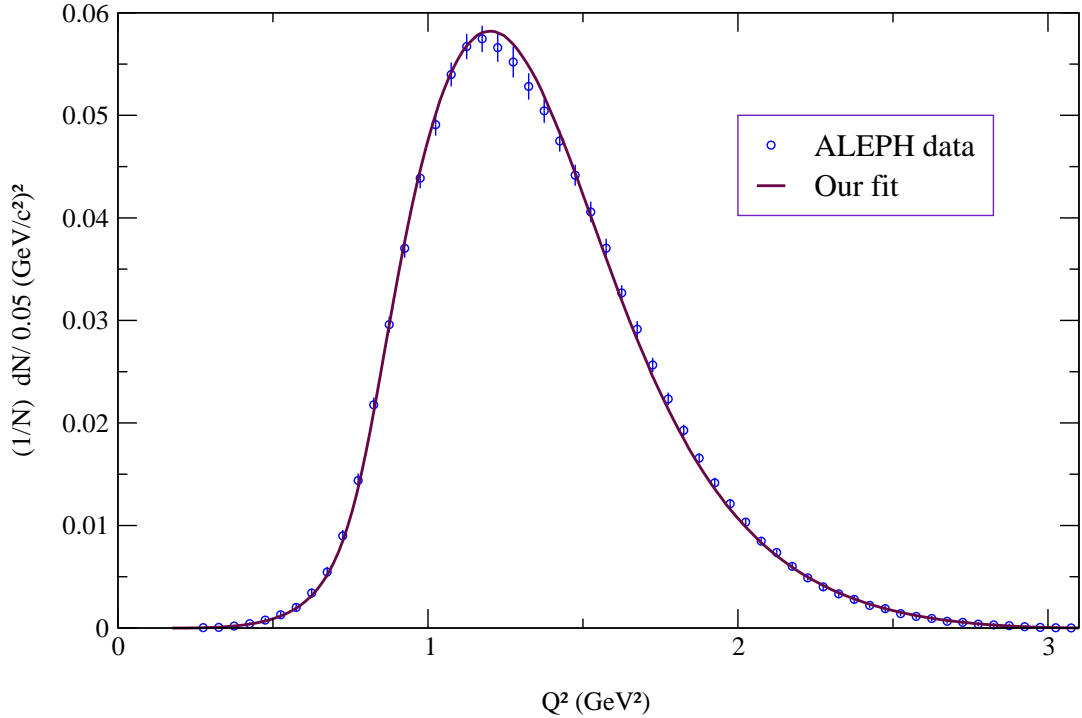}
\caption{\label{fig:sp} Fit to the ALEPH data \cite{Barate:1998uf}
for the normalized $\tau^- \rightarrow \pi^+ \pi^- \pi^- \nu_{\tau}$. }
\end{figure}

We have fitted the experimental values for the $\tpppc$ branching ratio
and normalized spectral function obtained by ALEPH \cite{Barate:1998uf} and 
we get a reasonable 
$\chi^2/d.o.f. = 64.5 / 52$. This is shown in Fig.~\ref{fig:sp}. 
Hence we get the axial--vector
$a_1(1260)$ parameters $M_A = (1.203 \pm 0.003) \, \mbox{GeV}$ and 
$\Gamma_{a_1}(M_A^2) = (0.48 \pm 0.02) \, \mbox{GeV}$, where the errors
are only statistical. We also obtain a value for the still unknown
combination of $\lambda_i$ couplings~: $\lambda_0 = 11.9 \pm 0.4$. 
However, as pointed out in Ref.~\cite{Cirigliano:2004ue}, this value
seems too large when additional QCD constraints are imposed.
The origin of the discrepancy could be the 
small sensitivity of the tau decay amplitude to this parameter, as it only
appears multiplied by the mass of the pion (\ref{eq:fq2}), together with an 
improvable implementation of the off-shell width of the $a_1(1260)$ resonance.
\par
\begin{figure}
  \includegraphics[angle=-90,scale=0.66,clip]{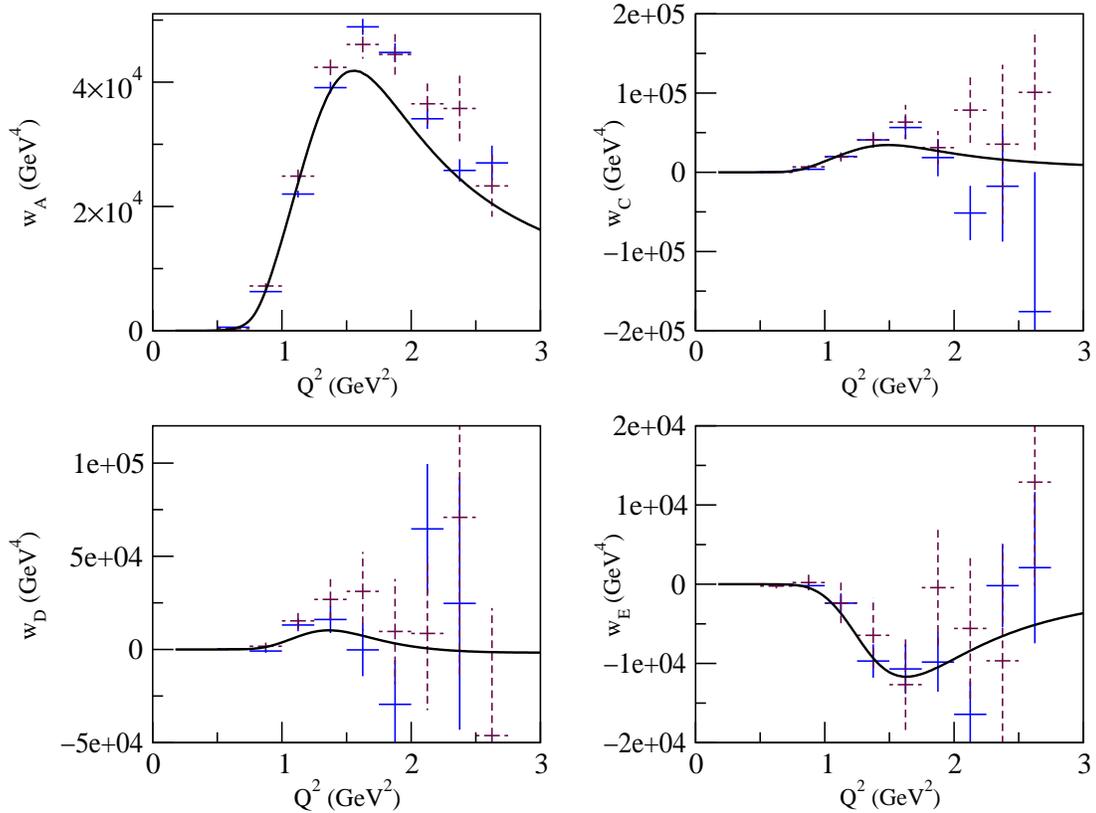}
  \caption{\label{fig:sf} Theoretical values for the $w_A$, $w_C$, $w_D$
and $w_E$ integrated structure functions in comparison with the experimental
data from CLEO-II (solid) and OPAL (dashed) \cite{Ackerstaff:1997dv}. }
\end{figure}

Ultimately we predict the integrated structure functions $w_A$, 
$w_C$, $w_D$ and $w_E$ 
\cite{Kuhn:1992nz}, that we compare with the experimental results for 
$\tpppn$ in Fig.~\ref{fig:sf}. In spite of the large errors the predictions
follow notably the depicted trend. As analysed in 
Ref.~\cite{GomezDumm:2003ku} a variance between the ALEPH data on one side
and the OPAL and CLEO-II on the other are at the origin of
the seemingly inconsistent result for $w_A$ in the high $Q^2$ region.
\par
In conclusion it can be inferred that, within the present experimental
errors and for
the studied observables, there is no evidence of relevant contributions 
in $\tppp$ decays beyond those of the $\rho(770)$ and $a_1(1260)$ resonances.
%


\begin{theacknowledgments}
J.P. wishes to thank Pietro Colangelo and Fulvia De Fazio for the
remarkable organization of the QCD@Work 2005 meeting in Conversano (Italy).
This work has been supported in part by
MEC (Spain) under grant FPA2004-00996, by Generalitat Valenciana
(Grants GRUPOS03/013, GV04B-594 and GV05/015) and by HPRN-CT2002-00311 
(EURIDICE).
\end{theacknowledgments}



\bibliographystyle{aipproc}   


\begin{thebibliography}{9}

\bibitem{Pich:1989pq}
  A.~Pich,
  ``QCD Tests From Tau Decay Data,''
{\it Talk given at Tau Charm Factory Workshop,
 Stanford, Calif., May 23-27, 1989} ;
  J.~H.~Kuhn and A.~Santamaria,
  Z.\ Phys.\ C {\bf 48} (1990) 445.
  
\bibitem{Achasov:1996gw}
  N.~N.~Achasov and A.~A.~Kozhevnikov,
  Phys.\ Rev.\ D {\bf 55} (1997) 2663
  [arXiv:hep-ph/9609216].

\bibitem{Portoles:2000sr}
  J.~Portol\'es,
  Nucl.\ Phys.\ Proc.\ Suppl.\  {\bf 98} (2001) 210
  [arXiv:hep-ph/0011303];
 
\bibitem{GomezDumm:2003ku}
  D.~G\'omez Dumm, A.~Pich and J.~Portol\'es,
  Phys.\ Rev.\ D {\bf 69} (2004) 073002
  [arXiv:hep-ph/0312183].
  
\bibitem{Portoles:2004vr}
  J.~Portol\'es,
  Nucl.\ Phys.\ Proc.\ Suppl.\  {\bf 144} (2005) 3
  [arXiv:hep-ph/0411333].
  
\bibitem{Weinberg:1978kz}
  S.~Weinberg,
  PhysicaA {\bf 96} (1979) 327;
  J.~Gasser and H.~Leutwyler,
  Annals Phys.\  {\bf 158} (1984) 142.

\bibitem{Colangelo:1996hs}
  G.~Colangelo, M.~Finkemeier and R.~Urech,
  Phys.\ Rev.\ D {\bf 54} (1996) 4403
  [arXiv:hep-ph/9604279].
  
\bibitem{Barate:1998uf}
  R.~Barate {\it et al.}  [ALEPH Collaboration],
  Eur.\ Phys.\ J.\ C {\bf 4} (1998) 409.

\bibitem{Ackerstaff:1997dv}
  K.~Ackerstaff {\it et al.}  [OPAL Collaboration],
  Z.\ Phys.\ C {\bf 75} (1997) 593;
  T.~E.~Browder {\it et al.}  [CLEO Collaboration],
  Phys.\ Rev.\ D {\bf 61} (2000) 052004
  [arXiv:hep-ex/9908030].

\bibitem{Ecker:1988te}
  G.~Ecker, J.~Gasser, A.~Pich and E.~de Rafael,
  Nucl.\ Phys.\ B {\bf 321} (1989) 311.
  
\bibitem{Ecker:1989yg}
  G.~Ecker, J.~Gasser, H.~Leutwyler, A.~Pich and E.~de Rafael,
  Phys.\ Lett.\ B {\bf 223} (1989) 425.

\bibitem{'tHooft:1974hx}
  G.~'t Hooft,
  Nucl.\ Phys.\ B {\bf 75} (1974) 461;
  E.~Witten,
  Nucl.\ Phys.\ B {\bf 160} (1979) 57.

\bibitem{Pich:2002xy}
  A.~Pich, in Proceedings of the Phenomenology of Large $N_C$ QCD, 
  edited by R.~Lebed (World Scientific, Singapore, 2002), p.~239,
  arXiv:hep-ph/0205030.
  
\bibitem{Kuhn:1992nz}
  J.~H.~Kuhn and E.~Mirkes,
  Z.\ Phys.\ C {\bf 56} (1992) 661
  [Erratum-ibid.\ C {\bf 67} (1995) 364].

\bibitem{GomezDumm:2000fz}
  D.~Gomez Dumm, A.~Pich and J.~Portol\'es,
  Phys.\ Rev.\ D {\bf 62} (2000) 054014
  [arXiv:hep-ph/0003320].
    
\bibitem{Floratos:1978jb}
  E.~G.~Floratos, S.~Narison and E.~de Rafael,
  Nucl.\ Phys.\ B {\bf 155} (1979) 115.
  
\bibitem{Cirigliano:2004ue}
  V.~Cirigliano, G.~Ecker, M.~Eidemuller, A.~Pich and J.~Portol\'es,
  Phys.\ Lett.\ B {\bf 596} (2004) 96
  [arXiv:hep-ph/0404004].
  	       
\end{thebibliography}



\end{document}